\documentclass[12pt]{extarticle}

\usepackage{lmodern} 
\usepackage[margin=1in]{geometry} 
\usepackage{setspace}
\usepackage{amsmath, amssymb, amsfonts, amsthm, bm}

\newtheorem{theorem}{Theorem}[section]

\newtheorem{remark}[theorem]{Remark}
\newtheorem{corollary}{Corollary}
\usepackage{graphicx, subcaption, tikz}
\usepackage{booktabs, multirow, threeparttable}
\usepackage{algorithm}
\usepackage{algpseudocode}

\usepackage[round,authoryear]{natbib}

\usepackage[colorlinks=true, linkcolor=blue, citecolor=blue, urlcolor=blue]{hyperref}

\usepackage{abstract}

\usepackage{CJKutf8}

\title{\bfseries Robust Clustered Federated Learning for Heterogeneous High-dimensional Data}

\author{
\large Changxin Yang$^{1}$, Zhongyi Zhu$^{1}$, Heng Lian$^{2}$\\[1ex]
\normalsize $^{1}$Department of Statistics and Data Science, Fudan University\\
\normalsize $^{2}$Department of Mathematics, City University of Hong Kong
}

\date{} 

\begin{document}

\maketitle
\begin{abstract}
\noindent 
Federated learning has attracted significant attention as a privacy-preserving framework for training personalised models on multi-source heterogeneous data. However, most existing approaches are unable to handle scenarios where subgroup structures coexist alongside within-group heterogeneity. In this paper, we propose a federated learning algorithm that addresses general heterogeneity through adaptive clustering. Specifically, our method partitions tasks into subgroups to address substantial between-group differences while enabling efficient information sharing among similar tasks within each group. Furthermore, we integrate the Huber loss and Iterative Hard Thresholding (IHT) to tackle the challenges of high dimensionality and heavy-tailed distributions. Theoretically, we establish convergence guarantees, derive non-asymptotic error bounds, and provide recovery guarantees for the latent cluster structure. Extensive simulation studies and real-data applications further demonstrate the effectiveness and adaptability of our approach.
\end{abstract}


\section{Introduction}
High-dimensional datasets from diverse sources, such as medical records or financial data from different regions, are increasingly common in real-world applications. These datasets often contain valuable shared information, and integrating them can significantly improve predictive accuracy, as demonstrated by \cite{10.1111/biom.12177}. However, such integration also poses several challenges. Firstly, direct data sharing is often infeasible due to privacy concerns and limited transmission capacity. Researchers from different data sources are typically restricted to sharing only summary statistics. Additionally, data from different sources may exhibit heterogeneity. For example, electronic health records from different hospitals may exhibit subgroup structures due to regional differences in patient populations. Simultaneously, even within the same subgroup, slight heterogeneity may persist due to variations in population composition, healthcare systems, or other social factors, leading to similar but non-identical models. Ignoring such heterogeneity may result in biased estimates and degraded predictive performance. Furthermore, the challenges of high dimensionality, combined with heavy-tailed distributions, add further complexity to the analysis. For instance, \cite{Wang02102015} shows that certain gene expression levels in microarray data display heavy-tailed behaviour, under which standard sparse least squares methods often perform unsatisfactorily.
While a broad range of studies have tackled subsets of these challenges, only a few methods are capable of addressing all aspects jointly.
\par Distributed optimisation is a common approach for managing data stored across multiple locations. Such methods are generally categorized into two types: centralised and decentralised. Centralised approaches, such as \cite{doi:10.1080/01621459.2018.1429274} and \cite{doi:10.1080/01621459.2021.1969238}, require a central server to aggregate information from all participating tasks and optimise an approximate global objective, thereby achieving higher efficiency. Conversely, decentralised methods, including \cite{Liu2022FastAR} and \cite{8988200}, employ graph-based network communication systems to avoid single points of failure and distribute computational costs across tasks. However, most existing approaches assume task homogeneity, which may not hold in practice.

\par To address heterogeneity in multi-source data, multitask learning and transfer learning are commonly employed. \cite{10.1214/23-AOS2319} proposed a multitask learning (ML) framework that estimates multiple heterogeneous tasks by adaptively clustering and extracting within-group information, and \cite{lam2022adaptivedatafusionmultitask-} extended this framework to the nonsmooth case. \cite{10.1111/rssb.12479} adopted a transfer learning approach that first computes an initial estimator using informative auxiliary samples, followed by a bias-correction step based on the target data, while \cite{doi:10.1080/01621459.2023.2184373} extended this approach to high-dimensional generalized linear models. However, the majority of existing work on both types of methods typically requires access to complete data from all tasks, which violates privacy constraints.

\par Personalised Federated Learning (FL) develops task-specific models to address heterogeneity while preserving privacy. Existing methods can be categorized based on the type of heterogeneity they handle. 
The first category assumes tasks are similar but not identical, treating them as a single group with within-group heterogeneity. A notable work is \cite{10.1093/biostatistics/kxv038}, which proposes a sparse meta-analysis for multiple studies but lacks theoretical guarantees. Similarly, \cite{doi:10.1080/01621459.2021.1904958} developed a one-shot method for high-dimensional meta-analysis with theoretical consistency. These approaches perform well with mild heterogeneity within the group, but struggle with significant variations between tasks from mixed groups.
The second category involves FL based on clustering, where heterogeneity arises from latent subgroup structures, assuming homogeneity within each group. For example, \cite{9832954} used the $k$-means algorithm for task clustering, while \cite{NEURIPS2021_82599a4e} employed the EM algorithm to identify group structures. However, both require pre-specified cluster numbers and lack sparsity-inducing regularisation in high-dimensional settings. In contrast, \cite{JMLR:v25:23-0059} and \cite{doi:10.1080/01621459.2024.2321652} propose sparse fused penalty-based frameworks that adaptively determine the number of clusters and capture grouping structures.
However, their theoretical guarantees depend on the assumption of homogeneity within the group. In the presence of within-group heterogeneity, the oracle estimator may no longer satisfy local optimality, and misclustering can significantly affect the accuracy of the estimation.

\par To handle high-dimensional data with heavy-tailed noise, sparse Huber regression is a powerful tool, as it only requires {low-order moments} to ensure consistency. \cite{doi:10.1080/01621459.2018.1543124} established non-asymptotic deviation bounds for $\ell_1$-penalized adaptive Huber regression, while \cite{10.1016/j.csda.2021.107419} and \cite{https://doi.org/10.1111/sjos.12723} extended adaptive Huber regression to distributed and transfer learning contexts. In this paper, we employ the IHT method to address high-dimensional Huber regression. Due to the inherent properties of the constraint $\ell_0$, IHT is particularly effective for variable selection. \cite{NIPS2014_218a0aef} and \cite{WANG202336} demonstrate that IHT methods achieve linear convergence for least squares regression and several non-smooth problems with $\ell_0$-constraints. Motivated by these observations, we extend the IHT framework to the Huber loss, offering a perspective distinct from $\ell_1$-based approaches for high-dimensional Huber regression.

\par This work proposes a robust clustered FL framework for high-dimensional regression that accommodates both within- and between-group heterogeneity. Specifically, for each task, we compute the gradient of the Huber loss and transmit it to the central server. The server applies the IHT method to update the sparse robust local estimators. It then uses pairwise $\ell_2$ distance penalties to reveal the underlying group structure, leveraging task similarities to improve estimation efficiency. This approach facilitates clustering and information extraction even in the presence of within-group heterogeneity, as suggested by \cite{10.1214/23-AOS2319}. Through alternating updates between local and central stages, each task achieves a personalised model informed by similar tasks while preserving privacy.

\par Compared with existing work, our approach makes several notable contributions. 
First, we develop novel algorithms for personalised FL and establish their linear convergence, ensuring both data privacy and computational efficiency. 
Second, our framework accommodates both within- and between-group heterogeneity through a pairwise $\ell_2$ group penalty, providing a more flexible approach for heterogeneous tasks. In particular, we first extend the clustering approach of \cite{10.1214/23-AOS2319} to the high-dimensional FL and establish theoretical guarantees for recovering the latent group structure. By leveraging task similarities within each group, our method enhances the performance of individual tasks. Third, we integrate the IHT method with the Huber loss to create a robust and sparse regression method. The IHT method allows control over the number of selected variables, which is particularly advantageous for scientific applications such as gene selection in biomedical research. The incorporation of the Huber loss further enhances robustness against outliers and heavy-tailed noise. Based on these strengths, we establish non-asymptotic deviation bounds and support recovery guarantees for the proposed estimators.

\par The remainder of the paper is organized as follows. Section \ref{s2} formulates the robust regression problem under a general heterogeneous framework and introduces the proposed algorithms based on the IHT method.
Section \ref{s4} provides theoretical analysis, establishing convergence guarantees and error bounds. We evaluate the proposed method using simulated data in Section~\ref{simu} and real-world data in Section~\ref{s6}.
 Finally, Section \ref{s7} concludes the paper with a summary and potential directions for future research.

\subsection{Notation}
We introduce some necessary notations. Let $c$ and $C$ denote positive constants independent of $n$ and $p$. For simplicity, we denote $[n]$ as the set $\{1, \dots, n\}$. The $\ell_q$ norm for $\boldsymbol{a} \in \mathbb{R}^p$ is defined as
$\|\boldsymbol{a}\|_q = \left(\sum_{j \in [p]} |a_j|^q\right)^{1/q},$
while $\|\boldsymbol{a}\|_0$ represents the number of nonzero entries in $\boldsymbol{a}$. Denote the support of $\boldsymbol{a}$ as $\operatorname{supp}(\boldsymbol{a})$.
For positive sequences $a_n$ and $b_n$, the notation $a_n = O(b_n)$ or $a_n \lesssim b_n$ indicates the existence of a constant $c$ such that $a_n / b_n \leq c$ for all $n \geq 1$. Similarly, $a_n = o(b_n)$ implies $a_n / b_n \to 0$, while $a_n \asymp b_n$ signifies that $a_n = O(b_n)$ and $b_n = O(a_n)$. The notation $a_n \ll b_n$ or $b_n \gg a_n$ is equivalent to $a_n = o(b_n)$. For a set $I$, we define $|I|$ as its cardinality. 

\section{Problem setup}\label{s2}

We consider a federated learning problem involving \( M \) heterogeneous tasks with balanced sample sizes. For each task \( m \), we observe a local dataset \(\mathcal{D}_m = \left\{\mathbf{x}_{m i}, y_{m i}\right\}_{i=1}^{n}\) generated from the linear model
$$
y_{mi} = \mathbf{x}_{mi}^{\top} \boldsymbol{\beta}_m^* + \epsilon_{mi}, \quad i = 1, \ldots, n,
$$
where \(\boldsymbol{\beta}_m^* \in \mathbb{R}^p\) is the parameter of primary interest for the \( m \)-th task. The independent random errors \( \epsilon_{mi} \) are mean-zero and satisfy \(\mathbb{E}\left(\left|\epsilon_{mi}\right|^{1+\delta}\right) < \infty \) for some \( 0<\delta \leq 1 \), which allows for the presence of heavy tails.
Throughout, we consider the high-dimensional regime where the dimension \( p \) is much larger than the number of observations, with  \( \left\|\boldsymbol{\beta}_m^*\right\|_0\leq s_0 \).

\par We adopt a general heterogeneity framework to enhance the ability to handle heterogeneity among tasks.
Specifically, we assume the existence of $K$ groups $\{\mathcal{G}_k\}_{k \in [K]}$, 
forming a non-overlapping partition of $[M]$, with associated group centres 
$\{\boldsymbol{\theta}_k^\star\}_{k \in [K]} \subseteq \mathbb{R}^p$. 
For the $k$-th group, 
$\mathcal{G}_k = \{m: z_m^\star = k\}$, where $z_m^\star$ denotes the latent cluster label of task $m$. 
The within-group heterogeneity is bounded by a constant $h$: 
$
\max_{m} \|\boldsymbol{\beta}_m^\star - \boldsymbol{\theta}_{z_m^\star}^\star\|_2 \leq h.
$
This framework recovers classical meta-analysis when $h \geq 0, K=1$ (\cite{doi:10.1080/01621459.2021.1904958}), 
and reduces to clustering-based FL when $h=0, K>1$ (\cite{9832954}). 
It thus extends the heterogeneity structures in most existing personalised FL methods. To account for within-group variability, we define the group support as
$
I_k^\star = \{j \in [p] : \exists m \in \mathcal{G}_k \text{ s.t. } \beta_{mj}^\star \neq 0\},
$
and assume $|I_k^\star| \leq q_0$, where $q_0$ reflects similarity of supports within each group. 
Between-group separation is measured by
$
\Delta \doteq \min_{k \neq k^\prime} \|\boldsymbol{\theta}_k^\star - \boldsymbol{\theta}_{k^\prime}^\star\|_2.
$
We further impose a balance condition: 
$\min_{k \in [K]} |\mathcal{G}_k| \geq c_0 M / K$ for some constant $c_0 > 0$.  
This framework was originally proposed in \cite{10.1214/23-AOS2319} for low-dimensional multi-task learning. The methodologies and theoretical results developed in this paper extend it for the first time to high-dimensional federated learning.

\par Our goal is to identify subgroup structures and extract information from group members to improve the estimation for individual tasks. 
Building on ideas from \cite{tian2024theory}, we propose a federated learning framework that employs an iterative algorithmic strategy. Given a total of \(T\) rounds, at each iteration \(t = 1, \dots, T\), every task computes its local gradient and sends it to the central server 
for updating the sparse local estimators \(\boldsymbol{\beta}_m^{(t)}\). 
Subsequently, the central server solves the following optimization problem:
\begin{equation}
    \begin{aligned}
\underset{
    \substack{
        \boldsymbol{\beta} \in \mathbb{R}^{p \times M},\;
        \boldsymbol{\theta} \in \mathbb{R}^{p \times K},\;
        \mathbf{z} \in [K]^M
    }
}{\arg \min} \;\;
& \sum_{m=1}^M \left( \frac{1}{2} \left\| \boldsymbol{\beta}_m - {\boldsymbol{\beta}}_m^{(t)} \right\|_2^2 
+ \lambda \left\| \boldsymbol{\beta}_m - \boldsymbol{\theta}_{z_m} \right\|_2 \right), \\
\text{subject to} \quad
& \|\boldsymbol{\beta}_m\|_0 \leq s, \quad \text{for all } m = 1, \dots, M, \label{ls1}
\end{aligned}
\end{equation}
{where \(\boldsymbol{\beta} = \left( \boldsymbol{\beta}_1, \ldots, \boldsymbol{\beta}_M \right)\), \(\mathbf{z} = (z_1, \dots, z_M)\), and \(\boldsymbol{\theta} = (\boldsymbol{\theta}_1, \dots, \boldsymbol{\theta}_K)\). \(K\) denotes the pre-specified number of groups, and \(s\) is a positive integer controlling the sparsity level. The penalty term promotes subgroup identification and information sharing across tasks, with $\lambda>0$ serving as the regularization parameter.
This strategy enables the adaptive identification of subgroup structures and facilitates information sharing among tasks within each group. 
After a few rounds of alternating local and central updates, the algorithm produces the final federated estimators. 
\par Our method offers notable advantages over existing approaches. Most clustering-based federated learning methods struggle in high-dimensional regimes with general heterogeneity. Sparse fused penalty–type approaches (\cite{doi:10.1080/01621459.2024.2321652, JMLR:v25:23-0059}) impose sparse penalties (e.g., SCAD, $\ell_1$) on pairwise distances among tasks to encourage clustering. However, in the presence of within-group heterogeneity, their performance deteriorates: the oracle estimator may fail to retain local optimality, and misclustering can substantially degrade estimation accuracy. \cite{9832954} proposed a $k$-means-based framework assuming identical models within each group, later extended by \cite{osti_10525022} to handle both within- and between-group heterogeneity without knowing the number of clusters. However, their method requires $p = o(n^2)$ and lacks variable selection, limiting applicability in sparse high dimensions. Our method extends the clustering framework of \cite{10.1214/23-AOS2319} to high-dimensional regimes. They showed that using $\ell_2$ distances allows for effective clustering and information extraction in the presence of within-group heterogeneity. However, the \(\ell_2\)-type group penalty can suffer from the curse of dimensionality. To our knowledge, there is no existing work that directly extends this to high-dimensional settings. To this end, we adopt the $\ell_0$-constrained surrogate loss in \eqref{ls1} based on sparse local estimators, and demonstrate that, under mild conditions, our method achieves accurate subgroup recovery and consistent estimation across tasks.

\section{ Methodology }\label{s3}
In this section, we first present the IHT method for high-dimensional Huber regression, and then extend it to the federated learning setting.
\subsection{IHT Method for Huber Regression}\label{subsec2}
In this section, we first present the IHT method for high-dimensional Huber regression. For each task \( m \), we optimise the local objective function 
\begin{equation*}
    H_\sigma^m(\boldsymbol{\beta}_m) \doteq \frac{1}{n} \sum_{i=1}^{n} H_\sigma \left( y_{m i} - \mathbf{x}_{m i}^{\top} \boldsymbol{\beta}_m \right) \quad \text{subject to} \quad \|\boldsymbol{\beta}_m\|_0 \leq s,
\end{equation*}
where
$
    H_\sigma(x) = \frac{x^2}{2} \cdot \mathbb{I}(|x| \leq \sigma) + \left(\sigma |x| - \frac{\sigma^2}{2}\right) \cdot \mathbb{I}(|x| > \sigma)
$
and \(\sigma\) is a robustification parameter that balances efficiency and robustness to heavy-tailed noise. Although the $\ell_0$ norm is well-suited for variable selection and offers good interpretability, 
developing efficient algorithms for $\ell_0$-constrained problems remains challenging. 
Several studies have investigated this issue (\cite{doi:10.1073/pnas.2014241117, xue2021constrained, NIPS2014_218a0aef, WANG202336}). 
Among them, the IHT procedure is particularly appealing due to its computational simplicity 
and has been shown to converge linearly.

\par Specifically, starting from an initial estimator \( \boldsymbol{\beta}^{(0)}_m \) (e.g., \( \boldsymbol{0} \in \mathbb{R}^p \)) for the \( m \)th task, the local IHT algorithm proceeds as
\begin{equation} \label{local}
    \boldsymbol{\beta}^{(t)}_m = P_{s}\left(\boldsymbol{\beta}_m^{(t-1)} - \eta \nabla H_\sigma^m\left(\boldsymbol{\beta}_m^{(t-1)}\right)\right), \quad t = 1, 2, \ldots,
\end{equation}
where \( \eta > 0 \) is a step size, \( \nabla H_\sigma^m(\cdot) \) denotes the gradient of the Huber loss, and \(P_s(\cdot)\) is a hard-thresholding projection operator that retains the \(s\) entries with the largest magnitudes in the input vector, setting all remaining entries to zero:
\begin{equation}
    \left(P_s\left(\boldsymbol{\alpha}\right)\right)_j = 
\begin{cases}
\alpha_j, & \text{if } j \text{ is among the top } s \text{ indices of } \{|\alpha_1|, \ldots, |\alpha_p|\}, \\
0, & \text{otherwise},
\end{cases}
\quad \text{for } \boldsymbol{\alpha} \in \mathbb{R}^p. \label{ps}
\end{equation}
 \begin{center}  
\begin{minipage}{\linewidth}  
\begin{algorithm}[H]
\small
\caption{Local IHT Algorithm for Huber Regression}\label{localIHt}
\begin{algorithmic}[1]
\Require Initialise $\{{\boldsymbol{\beta}}^{(0)}_m\}_{m \in [M]}$, data $\left\{ \mathbf{x}_{mi}, y_{mi} \right\}_{i \in [n], m \in [M]}$, iteration number $T$, step size $\eta$, sparsity $s$; 
 \For {task $m = 1$ to $M$}
   \For {$t = 1$ to $T$}
        \State Update $\boldsymbol{\beta}^{(t)}_m = P_{s}\left(\boldsymbol{\beta}_m^{(t-1)} - \eta \nabla H_\sigma^m\left(\boldsymbol{\beta}_m^{(t-1)}\right)\right)$.
    \EndFor
\EndFor
\State Final estimators $\{{{\boldsymbol{\beta}}}^{(T)}_m\}_{m \in [M]}$
    \end{algorithmic}
\end{algorithm}
\end{minipage}
\end{center}
 \par If each task is updated independently, the resulting estimator is referred to as the \textit{local IHT estimator}, as detailed in Algorithm~\ref{localIHt}. Theorem~\ref{teolocal} establishes that the local IHT estimator achieves nearly minimax-optimal performance under mild conditions. Supplementary Section~C.2 further shows its superior variable selection over $\ell_1$-penalized Huber regression (\cite{doi:10.1080/01621459.2018.1543124}), especially when signals are sparse and relatively strong. 

\par However, given the assumed similarity among certain tasks, a more effective approach is to transmit the local information to the central server. This allows for the aggregation of relevant information and further enhancement of estimation efficiency. 
\subsection{Federated IHT for Huber Regression \label{FLIHT}}

\par In this section, we propose a Federated IHT algorithm for heterogeneous task estimation. The algorithm iteratively communicates between the central server and local tasks, with each round consisting of two main steps. First, each local site computes its gradient and sends it to the central server. Second, the central server updates the sparse local estimators \(\boldsymbol{\beta}_m^{(t)}\) as in \eqref{gp} and implement a two-step optimization procedure for \eqref{ls1}:
\begin{itemize}
    \item \textbf{Step 1:} Solve the unconstrained optimization problem for 
$\{\tilde{\boldsymbol{\beta}}_m^{(t)}, \hat{z}_m^{(t)}\}_{m \in [M]}$ 
and $\{\hat{\boldsymbol{\theta}}_k^{(t)}\}_{k \in [K]}$:
     \begin{equation}\label{cu}
        \begin{aligned}
           \operatorname*{argmin}_{
                \substack{
                    \boldsymbol{\beta} \in \mathbb{R}^{p \times M} \\
                    \boldsymbol{\theta} \in \mathbb{R}^{p \times K}, \mathbf{z} \in [K]^M
                }
            }
        \mathcal{L}(\boldsymbol{\beta}, \boldsymbol{\theta}, \mathbf{z};\{\boldsymbol{\beta}_m^{(t)}\}_{m\in[M]})\doteq
                \sum_{m=1}^M 
                    \tfrac{1}{2}\left\|\boldsymbol{\beta}_m - {\boldsymbol{\beta}}_m^{(t)}\right\|_2^2 + 
                    \lambda \left\|\boldsymbol{\beta}_m - \boldsymbol{\theta}_{z_m}\right\|_2
        \end{aligned} 
\end{equation}

{where \(\boldsymbol{\beta} = \left( \boldsymbol{\beta}_1, \ldots, \boldsymbol{\beta}_M \right)\), \(\mathbf{z} = (z_1, \dots, z_M)\), and \(\boldsymbol{\theta} = (\boldsymbol{\theta}_1, \dots, \boldsymbol{\theta}_K)\)}
     \item \textbf{Step 2:} Perform sparse projection by solving:
    \begin{equation}
        \hat{\boldsymbol{\beta}}_m^{(t)} = P_s(\tilde{\boldsymbol{\beta}}_m^{(t)}) = \underset{\|\boldsymbol{\beta}\|_0 \leq s}{\arg\min} \|\boldsymbol{\beta} - \tilde{\boldsymbol{\beta}}_m^{(t)}\|_2,
    \quad \text{for } m \in [M]. \label{localps}
    \end{equation}
\end{itemize} The first step aims to capture the relatedness among tasks, while the second step approximates the globally optimal solution using an \( s \)-sparse vector. To overcome the curse of dimensionality associated with the $\ell_2$
  group penalty, we apply two distinct projections at each round of the optimization process: first, a sparse projection in \eqref{gp} reduces the impact of noisy features for high-dimensional clustering; second, a projection in \eqref{localps} removes within-group support heterogeneity, ensuring that personalised estimators are $s$-sparse.

\par To motivate the need for the first projection, a critical observation is that the latent group structure depends only on a specific subset of features, e.g., \( j \in I_k^* \). Numerous uninformative features will make the pairwise Euclidean distances in \eqref{ls1} less informative and diminish the accuracy of the final clustering results. To address this challenge, machine learning algorithms usually employ various strategies to select or discard features. In sparse \(k\)-means clustering, feature-weighting and feature-ranking techniques are widely used in \cite{pmlr-v108-chakraborty20a, NEURIPS2020_735ddec1, 9309172, pmlr-v206-chakraborty23a}.
These methods share the common principle of focusing only on distances computed from important features for clustering, but differ in how these features are identified.
 
\par Inspired by sparse \(k\)-means clustering, we apply a projection operator to the local estimator for dimensionality reduction and variable selection. In the previous section, we employed the traditional projection operator \(P_s(\cdot)\) to perform dimensionality reduction and variable selection after each gradient descent update. However, the feature ranking in \eqref{ps} relies solely on local samples from each task. Ignoring potentially useful information from tasks within the same subgroup may impede the identification of clustering-relevant features, particularly when the local sample size is limited. More importantly, if a task erroneously discards certain important features, the information from that task regarding those features is permanently lost, as the corresponding values are set to zero. This, in turn, reduces the effectiveness of the solution to \eqref{ls1}, even if the task is correctly classified. To address these issues, we propose a group projection operator that facilitates the construction of more effective local estimators by incorporating group-level information.
}
\par  Specifically, we replace the local update in \eqref{local} with a group projection step:
\begin{equation}
    \boldsymbol{\beta}^{(t)}_{m} = P_{q}\left(\hat{\boldsymbol{\beta}}^{(t-1)}_{m} - \eta \nabla H_\sigma^{m}\left(\hat{\boldsymbol{\beta}}^{(t-1)}_{m}\right), \hat{\mathcal{G}}_k^{(t-1)}\right), \quad \text{for } m \in \hat{\mathcal{G}}_k^{(t-1)}, \label{gp}
\end{equation}
where \(\hat{\mathcal{G}}_k^{(t-1)} = \{m : \hat{z}_m^{(t-1)} = k\}\), and \(P_q(\cdot, \mathcal{G})\) denotes the group projection operator that retains only the coordinates corresponding to the top-\(q\) aggregated signals:
\begin{equation}
    (P_{q}(\boldsymbol{\alpha}_m, \hat{\mathcal{G}}_k))_j =
\begin{cases}
\alpha_{mj}, & \text{if } j \text{ is among the top } q \text{ of } \left\{ \left| \sum_{m \in \hat{\mathcal{G}}_k} \alpha_{mj} \right| \right\}_{j=1}^p. \\
0, & \text{otherwise},
\end{cases} \label{gprk}
\end{equation}
Here, \(q\) denotes the target sparsity level for the group support in practice. The ranking in \eqref{gp} is based on 
\(\sum_{m \in \hat{\mathcal{G}}_k^{(t-1)}} \hat{\boldsymbol{\beta}}^{(t-1)}_{m} - \eta \nabla H_\sigma^{m}\left(\hat{\boldsymbol{\beta}}^{(t-1)}_{m}\right)\),
which aggregates estimators within the group to identify important features. For a noise feature \(j \notin I_k^*\), it is less likely to be retained, as the aggregation causes the average of its estimates to converge to zero more rapidly. The set \(\hat{\mathcal{G}}_k^{(t-1)}\) plays a role analogous to the informative set in transfer learning frameworks (\cite{10.1111/rssb.12479}). Numerical results in Supplementary Sections~C.3 and C.4 demonstrate that the proposed group projection operator enhances both clustering accuracy and estimation efficiency compared with traditional operators or no projection in high-dimensional settings.
\par Now we focus on solving \eqref{cu}. Notably, the objective function does not have a closed-form solution and requires an exhaustive search across all possible classifications. To enable efficient implementation, we propose an iterative algorithm based on \(k\)-means. We start by reparameterizing the problem as \(\boldsymbol{\beta}_m = \boldsymbol{\theta}_{z_m} + \boldsymbol{\Delta}_m\), and then minimize the following objective:
\[
\sum_{m=1}^M \left(\frac{1}{2} \left\|\boldsymbol{\theta}_{z_m} + \boldsymbol{\Delta}_m - \boldsymbol{\beta}_m^{(t)}\right\|_2^2 + \lambda \left\|\boldsymbol{\Delta}_m\right\|_2\right)\,.
\]
The problem is solved using the following alternating optimization steps until convergence, which is defined by reaching a maximum number of iterations or when consecutive updates differ by less than a specified threshold:
\begin{itemize}
    \item \textbf{1. Initialization:}
    \begin{itemize}
    \item Set \(\boldsymbol{\Delta}_m = \mathbf{0}\) for all \(m \in [M]\). Apply the \(k\)-means algorithm to the local estimators \(\{\boldsymbol{\beta}_m^{(t)}\}_{m \in [M]}\) and obtain the corresponding cluster centers \(\{\tilde{\boldsymbol{\theta}}_k\}_{k \in [K]}\). These cluster centers are then transmitted to the local tasks, where each task determines its initial cluster assignment by solving
\begin{equation}
    z_m = \underset{z \in [K]}{\arg \min} \; H^m_\sigma(\tilde{\boldsymbol{\theta}}_z), \quad \text{for all } m \in [M]. \label{initial}
\end{equation}
    \end{itemize}
 \item \textbf{2. Update \(\boldsymbol{\theta}\) and \(\mathbf{z}\) while fixing \(\{\boldsymbol{\Delta}_m\}_{m \in [M]}\) using \(k\)-means:}
    \begin{itemize}
         \item Update \(\boldsymbol{\theta}\) while fixing \(\mathbf{z}\):
        \begin{equation}\label{theta}
            \boldsymbol{\theta}_{k} = \frac{\sum_{z_m = k} (\boldsymbol{\beta}_m^{(t)} - \boldsymbol{\Delta}_m)}{\sum_{m=1}^M I(z_m = k)}, \quad \text{for } k \in [K].
        \end{equation}
        \item {Update the group assignment}:
        \begin{equation}\label{z}
            z_m = \underset{z \in [K]}{\arg \min} \; \frac{1}{2} \left\|\boldsymbol{\theta}_{z} + \boldsymbol{\Delta}_m - \boldsymbol{\beta}_m^{(t)}\right\|_2^2, \quad \text{for } m \in [M].
        \end{equation}
    \end{itemize}
    \item \textbf{3. Update \(\{\boldsymbol{\Delta}_m\}_{m \in [M]}\) while fixing \(\boldsymbol{\theta}\) and \(\mathbf{z}\):}
    \begin{itemize}
        \item Perform the proximal update:
        \begin{equation}\label{delta}
            \boldsymbol{\Delta}_m^{\text{new}} = \operatorname{prox}_{\eta_1 \lambda}\left(
            \boldsymbol{\Delta}_m - \eta_1\left(\boldsymbol{\theta}_{z_m} + \boldsymbol{\Delta}_m - \boldsymbol{\beta}_m^{(t)}\right)\right),
        \end{equation}
        where \(\eta_1\) is the proximal map step size, and \(\operatorname{prox}_c(\boldsymbol{x}) = \left(1 - \frac{c}{\|\boldsymbol{x}\|_2}\right)_{+} \boldsymbol{x}\) denotes the proximal operator. Then set \(\boldsymbol{\Delta}_m = \boldsymbol{\Delta}_m^{\text{new}}\).
        \item Repeat the proximal update until convergence. This iterative procedure approximates the solution to the following optimization problem:
        \[
        \arg\min_{\boldsymbol{\Delta} \in \mathbb{R}^p} 
        \frac{1}{2} \left\| \boldsymbol{\theta}_{z_m} + \boldsymbol{\Delta} - \boldsymbol{\beta}_m^{(t)} \right\|_2^2 + \lambda \|\boldsymbol{\Delta}\|_2,
        \quad \text{for } m \in [M].
        \]
    \end{itemize}
    \item \textbf{4. Iteration:} Repeat Steps 2 and 3 until convergence.
\end{itemize}
 \par The basic idea of this alternating optimization framework is inspired by \cite{10.1214/23-AOS2319}. A key distinction is that we explicitly enforce the sparsity of \(\boldsymbol{\beta}_m^{(t)}\) via a projection operator at each iteration, thereby avoiding the curse of dimensionality. The numerical results show that our method exhibits reliable clustering performance even in challenging high-dimensional regimes. 
\begin{remark}
It is well known that the performance of \(k\)-means is sensitive to initialization. To improve stability, we do not directly adopt the \(k\)-means clustering results as initial values. Instead, the initial cluster centers are sent back to local tasks for reassignment using local data by \eqref{initial}, thereby enhancing initialization quality by leveraging richer local information---at the cost of one additional communication round.  Supplementary Section~C.1 compares this scheme with naive $k$-means initialization, highlighting its advantages.
\end{remark}
\begin{figure}[ht!]
  \centering
  \includegraphics[width=0.6\textwidth, height=0.25\textheight]{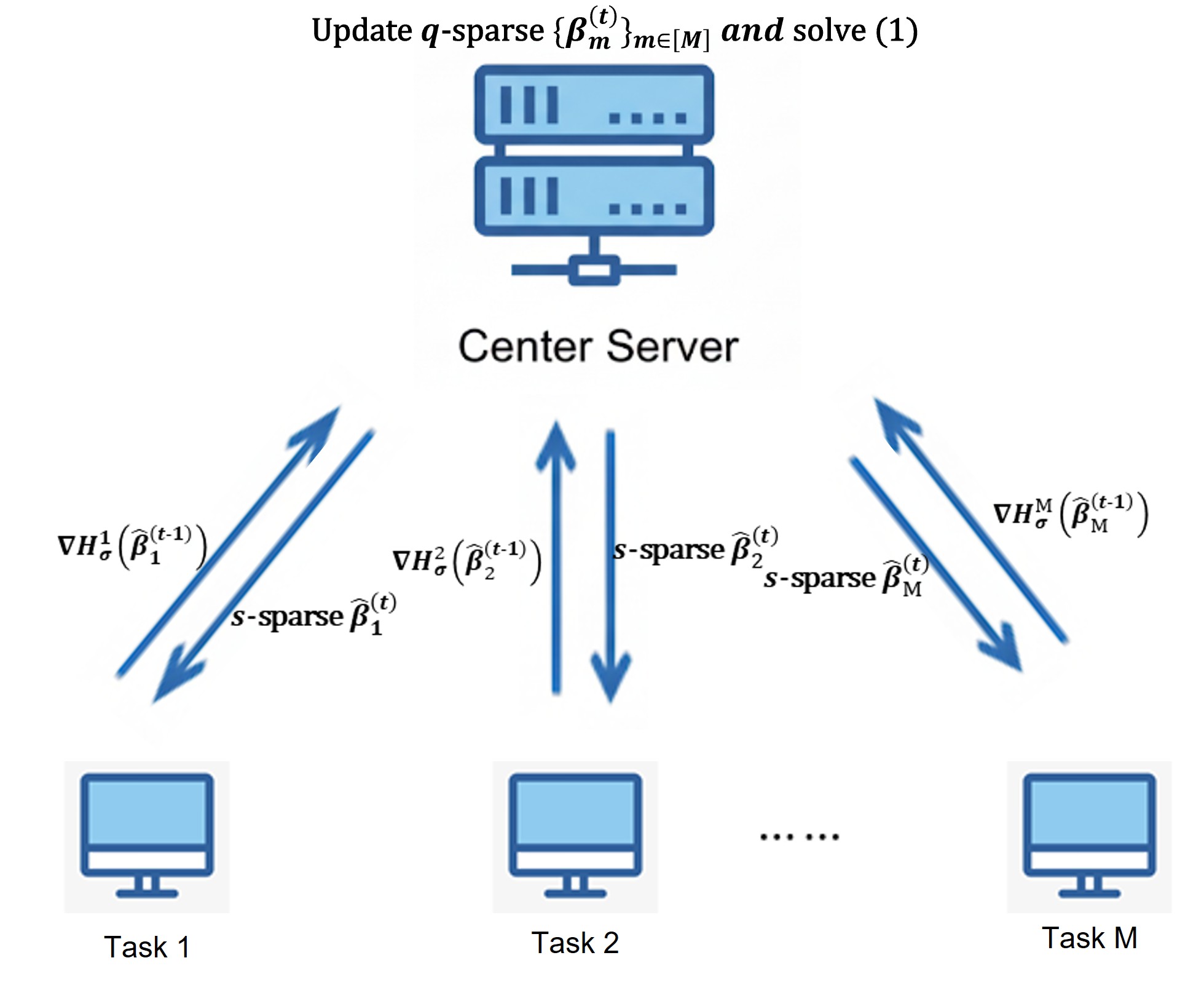}
  \caption{At the $t$-th iteration, each local task transmits the gradient to the central server, 
which first applies the group projection operator and then solves \eqref{ls1} to obtain the sparse federated estimators.}
\label{ar1}
\end{figure}
\par The overall algorithmic framework is summarized in Algorithm~\ref{fl}, and Figure~\ref{ar1} illustrates the architecture and communication scheme. As shown in Section~\ref{s4}, under appropriate heterogeneity, the federated error bound is substantially smaller than the local error bound. 
\begin{center}
\begin{minipage}{\linewidth}
\begin{algorithm}[H]
\small
\caption{{Federated IHT Algorithm with Group Projection}}\label{fl}
\begin{algorithmic}[1]
\Require Initialise $\{\hat{\boldsymbol{\beta}}^{(0)}_m\}_{m \in [M]}$ and $\{\hat{z}_m^{(0)}\}_{m \in [M]}$; $\{(\mathbf{x}_{mi}, y_{mi})\}_{i \in [n],\, m \in [M]}$; number of iterations $T, T_1$; step size $\eta$ and proximal map step size $\eta_1$; sparsity levels $s$ and $q$; number of groups $K$; group penalty parameter $\lambda$;
\For{\(t = 1\) to \(T\)}
    \State \textbf{Local Update:} For each task \(m = 1, \dots, M\), calculate \(\nabla H_\sigma^m(\hat{\boldsymbol{\beta}}^{(t-1)}_m)\).
    \State \textbf{Central Update:}
        \State Update the local estimators $\{\boldsymbol{\beta}_m^{(t)}\}_{m\in[M]}$ with \eqref{gp}.
    \State Update $\{\tilde{\boldsymbol{\beta}}_m^{(t)}, \hat{z}_m^{(t)}\}_{m \in [M]}$ 
and $\{\hat{\boldsymbol{\theta}}_k^{(t)}\}_{k \in [K]}$ by solving \eqref{cu}.    
    \State Update federated estimators \(\hat{\boldsymbol{\beta}}^{(t)}_m = P_{s}\left(\tilde{\boldsymbol{\beta}}^{(t)}_m\right)\).
\EndFor
\State Final federated estimators: \(\{\hat{\boldsymbol{\beta}}^{(T)}_m,\hat{z}_m^{(T)}\}_{m\in [M]}, \{\hat{\boldsymbol{\theta}}^{(t)}_k\}_{k \in [K]}\).
\end{algorithmic}
\end{algorithm}
\end{minipage}
\end{center}

\par \textbf{Tuning Parameter Selection}: The Huber loss parameter \(\sigma_m\) for each task \(m\) can be selected using its local data based on a variant of Lepski’s method, as proposed in \cite{doi:10.1080/01621459.2018.1543124}. We then set \(\sigma = \max_{m \in [M]} \sigma_m\). The local step size \(\eta\) in Algorithm~\ref{localIHt} and Algorithm~\ref{fl} can be chosen from a grid of candidate values by minimizing the local Huber loss for each task. Other hyperparameters—including \(\eta_1\), \(s\), \(q\), \(\lambda\), and \(K\)—are chosen by minimizing the following criterion based on training data:
\begin{equation}
     \frac{1}{Mn}\sum_{m=1}^M \sum_{i=1}^n H_\sigma\left(y_{mi} - \mathbf{x}_{mi}^\top \widehat{\boldsymbol{\beta}}_m^{(T)}\right) + \frac{\log p}{n}(C_1s+C_2K), \label{selec}
\end{equation}
where \(C_1, C_2\) are positive constants. This selection rule resembles the classical BIC criterion: the first term captures the predictive accuracy, while the last term penalizes model complexity.

\section{Theory \label{s4}}
\par In this section, we present convergence analysis for the proposed estimators. The proofs are provided in the supplementary material. We denote  
$
\mathcal{G}_{\min} \doteq \min_{k \in [K]} |\mathcal{G}_k|, \quad \text{and} \quad N_{\min} \doteq \mathcal{G}_{\min} n,
$
representing the minimal group size, and the minimal group sample size, respectively. Let \( e_0 \doteq \max_m \| \hat{\boldsymbol{\beta}}_m^{(0)} - \boldsymbol{\beta}_m^* \|_2 \) be the initial estimation error, and \(\beta_{\min} \doteq \min_{m\in[M],j \in [p]:\, \beta_{mj}^* \neq 0} |\beta_{mj}^*|\) the minimal signal strength. 
Moreover, we assume that samples are independent across different tasks and impose the following conditions:
\begin{itemize}
    \item \textbf{Assumption 1.} For any tasks \( m\in [M]\),  the covariate vectors \( \mathbf{x}_{mi} \in \mathbb{R}^p \) are i.i.d. within each task and drawn from a sub-Gaussian random vector \( \mathbf{x} \) with covariance \( \boldsymbol{\Sigma}_m = \mathbb{E}\left(\mathbf{x}_{mi} \mathbf{x}_{mi}^{\mathrm{T}}\right) \), i.e.,
\[
\mathbb{P}(|\langle \boldsymbol{u}, \mathbf{x}\rangle| \geq t) \leq 2 \exp \left(-t^2 \|\boldsymbol{u}\|_2^2 / A_0^2\right),
\]
for all \( t \in \mathbb{R} \) and \( \boldsymbol{u} \in \mathbb{R}^p \), where \( A_0 > 0 \) is a constant.
 \item \textbf{Assumption 2.}  For any task \( m \in [M] \), the regression errors \(\varepsilon_{mi}\) are i.i.d. within each task and satisfy \(\mathbb{E}\left(\varepsilon_{mi} \right) = 0\) and \(v_\delta \doteq \mathbb{E}\left(\left|\varepsilon_{mi}\right|^{1+\delta}\right) < \infty\) for some \(0 < \delta \leq 1\).
\item \textbf{Assumption 3}  
There exist constants \(0 < c_l \leq c_u < \infty\) such that for all unit vectors \(\mathbf{a}\) and all \(m \in [M]\),  
\begin{equation}
    c_l \leq \mathbf{a}^T \boldsymbol{\Sigma}_m \mathbf{a} \leq c_u. \label{a31}
\end{equation}

\end{itemize}
Assumptions~1--3 impose relatively mild conditions.  
Assumption~1 is a conventional requirement on random design matrices in high-dimensional settings.  
Assumption~2 allows for heavy-tailed distributions with a bounded \((1+\delta)\)-th moment and is commonly used in the analysis of Huber regression, seen in \cite{doi:10.1080/01621459.2018.1543124,doi:10.1080/01621459.2024.2321652}.  
Assumption~3 requires common eigenvalue bounds for all covariance matrices \(\{\boldsymbol{\Sigma}_m\}_{m \in [M]}\), which is standard in the literature \cite{doi:10.1080/01621459.2024.2321652,he2024transfusion,zhang2025transfer}.  
Additionally, we
comment that our theoretical analysis remains valid under a weaker restricted eigenvalue condition, as in \cite{NIPS2014_218a0aef,WANG202336}. For notational simplicity, we denote 
\(c_1=\left(\tfrac{2c_u - c_l}{2c_l}\right)^2\).

\subsection{Convergence of Local IHT Estimator}
\par We first present the convergence rate of independent task learning estimates. It helps to illustrate the advantages of the IHT method and provides motivation for our assumptions. 
 \begin{theorem}[Local IHT estimator] \label{teolocal} Suppose Assumptions 1, 2, and 3 hold for the $m$-th task. Let 
 $$
\sigma \asymp \left(v_\delta n / \log(p)\right)^{1 /(1+\delta)}, \quad \eta = \frac{4}{c_l + 2c_u},\quad  \min\left\{c_1,1\right\}\leq \frac{s}{s_0}< \infty.
 $$
 If $\sqrt{s_0}\left(\frac{\log (p )}{n}\right)^{\delta / (1+\delta)}=o(1)$ and
the initial error $e_0=o(\sigma)$, then, with probability at least \(1 - p^{-c}\) for some positive constant \(c\), and for all \(T \geq 1\), the local IHT estimators obtained by Algorithm~\ref{localIHt} satisfy:
\[
\begin{aligned}
 \left\|{\boldsymbol{\beta}}^{(T)}_m - \boldsymbol{\beta}^*_m\right\|_2 
\lesssim & \underbrace{\kappa_0^T}_{\text{iterative error}} + \underbrace{\sqrt{s_0}\left(\frac{\log (p )}{n}\right)^{\delta / (1+\delta)}}_{\text{local convergence rate}},
\end{aligned}
\]
where \( \kappa_0=\left(1 + \sqrt{\frac{s_0}{s}}\right) \frac{c_u - \frac{c_u}{2}}{c_u + \frac{c_u}{2}} + C \sqrt{\frac{s_0 \log(p)}{ n}}<1 \) for some positive constat $C$.
\end{theorem}

\par 
Theorem~\ref{teolocal} establishes the \textit{linear convergence} of our algorithm for Huber regression, inheriting the efficiency of previous IHT studies~\cite{NIPS2014_218a0aef,WANG202336}. Our requirements on the sparsity level $s$ and local sample size ensure that $\kappa_0 < 1$ holds. Consequently, $T = O\left(\log\left(e_0/\sqrt{s_0}\left(\tfrac{\log p}{n}\right)^{\delta/(1+\delta)}\right)\right)$ iterations suffice for our $s$-sparse estimator to achieve the rates:
\begin{equation}
\|\boldsymbol{\beta}_m^{(T)} - \boldsymbol{\beta}_m^*\|_2 
\lesssim \sqrt{s_0}\Bigl(\frac{\log p}{n}\Bigr)^{\frac{\delta}{1+\delta}}, \quad
\|\boldsymbol{\beta}_m^{(T)} - \boldsymbol{\beta}_m^*\|_1 
\lesssim s_0\Bigl(\frac{\log p}{n}\Bigr)^{\frac{\delta}{1+\delta}}.
\label{ll12}
\end{equation}
Similar bounds were obtained in \cite{doi:10.1080/01621459.2018.1543124, https://doi.org/10.1111/sjos.12723} for Lasso-penalized Huber regression. Our analysis requires an additional condition on the initial estimator,
$
e_0 = o(\sigma),
$
which is mild since $\sigma \asymp (n/\log p)^{1/(1+\delta)} \to \infty$. Moreover, IHT guarantees an \(s\)-sparse estimator, limiting selected variables to $O(s_0)$. In contrast, \cite{doi:10.1080/01621459.2018.1543124} provides no support recovery guarantees. Under suitable signal strength, the following corollary ensures recovery of all nonzero components of \(\boldsymbol{\beta}^*_m\), though some false positives may persist.
\begin{corollary}\label{cor1}
Under the assumptions of Theorem~\ref{teolocal}, if
$
\sqrt{s_0}\left(\tfrac{\log p}{n}\right)^{\delta/(1+\delta)}
\ll \beta_{min},
$ then after
$
T = O\left(\log\left(e_0/\sqrt{s_0}\left(\tfrac{\log p}{n}\right)^{\delta/(1+\delta)}\right)\right)
$ iterations,
with probability at least \(1 - p^{-c}\) for some constant \(c > 0\), the support of \(\boldsymbol{\beta}_m^{(T)}\) contains that of \(\boldsymbol{\beta}_m^*\), and the number of falsely selected variables is at most \(s - s_0\).
\end{corollary}
The signal strength requirements are standard in the literature, seen in \cite{10.1214/13-AOS1198, doi:10.1080/07350015.2023.2182309}. Notably, selecting $s$ closer to $s_0$ improves the variable selection performance of the IHT estimator in this case. Our analysis shows that the requirements on $\eta$ and $s$ are given by
\begin{equation}
    \left(1 + \sqrt{\tfrac{s_0}{s}}\right) 
    \max \Bigl\{\bigl|1 - \eta \tfrac{c_l}{2}\bigr|, \, \bigl|1 - \eta c_u\bigr|\Bigr\} < 1.  
    \label{vs}
\end{equation}
In particular, if the condition number of $\boldsymbol{\Sigma}_m$ is sufficiently well-behaved such that \eqref{vs} holds with $s = s_0$ and a reasonable choice of $\eta$, then the true support can be recovered without false positives. 
\subsection{Convergence of Federated IHT Estimator \label{orest}}
\par In the following, we analyse the estimator from the federated learning algorithms. 
\begin{itemize}
\item \textbf{Assumption 4.} 
There exists a positive constant $C_{\min} > 0$ such that for each group $k \in [K]$ 
and each $j \in I_k^*$,
$$
\frac{\left|\sum_{m \in \mathcal{G}_k} \beta_{mj}^*\right|}{|\mathcal{G}_k|} 
\geq C_{\min} \cdot \max_{m \in \mathcal{G}_k} |\beta_{mj}^*|.
$$
\item \textbf{Assumption 5.}
The within-group heterogeneity $h$ and the between-group separation $\Delta$ satisfy
\[\max\left\{Kh,\, K|\mathcal{G}_{\min}|^{1/(1+\delta)} \sqrt{q_0} 
\left(\frac{\log(pM)}{n}\right)^{\delta/(1+\delta)}\right\} \ll \Delta.
\]
\end{itemize}
Assumption~4 requires similar sign patterns and comparable signal strengths within groups. For instance, if $0<\beta_{m_1 j}^* \leq \cdots \leq \beta_{m_{\left|\mathcal{G}_k\right|} j}^*$ for each $j \in I_k^*$,  it is sufficient to require the existence of a constant $C_{min}$ such that
$
0<C_{\min } \leq \min _{j \in I_k^*} \frac{\beta_{m_1 j}^*}{\beta_{m_{\left|\mathcal{G}_k\right|}}^*}.
$
The constant $C_{\min}$ characterises the extent to which the strongest signal can be amplified through group aggregation.
Assumption 5 serves as a group structure identifiability condition, commonly adopted in clustering-based methods. The factor \(|\mathcal{G}_{\min}|^{1/(1+\delta)}\) arises from the larger robustification parameter \(\sigma \asymp \left( \frac{v_\delta N_{\min}}{ \log(pM) } \right)^{1/(1+\delta)}\) required in federated estimation, and vanishes under the standard squared loss.
When focusing on our model with balanced group sizes and $\delta = 1$, $h = 0$, and $|\cup_{k\in[K]} \operatorname{supp}(\boldsymbol{\theta}_k)| \asymp O(Ks_0)$,
\cite{JMLR:v25:23-0059} requires a smaller $\Delta$ for subgroup identification. However, it additionally assumes \(\frac{(Ks_0)^4 \log(p)}{n} = o(1)\), imposing a much stricter requirement on the local sample size to ensure asymptotic convergence. Meanwhile, \cite{doi:10.1080/01621459.2024.2321652} considers a feature-wise group structure and imposes an identifiability condition: 
$
K\sqrt{\frac{Ks_0 \log(p)}{n|\mathcal{G}_{\min}|}} \lesssim \min_{ k \neq k',\,j\in  I_{k^\prime}^*\cup I_k^* } \left| \theta_{k j}^* - \theta_{k' j}^* \right|,
$
 which further implies
$
K\sqrt{\frac{Ks_0^2 \log(p)}{n|\mathcal{G}_{\min}|}} \lesssim \Delta.
$
This makes their clustering approach more flexible, but potentially more restrictive than ours when $|\mathcal{G}_{\min}|^2 = o(Ks_0)$. Therefore, our condition characterises a trade-off between the local sample size and the separation level \(\Delta\), while allowing for a nonzero heterogeneity level \(h > 0\).

\par We first consider an oracle setting where the true clustering structure \(\{\mathcal{G}_k\}_{k \in [K]}\) is known, in order to illustrate the estimation efficiency under ideal cases. In this case, the objective function in \eqref{cu} simplifies to the following optimisation problem:
\begin{equation}
\begin{aligned}
      &\arg \min_{\boldsymbol{\beta}\in R^{|\mathcal{G}_k|\times p}} \left\{ \sum_{m \in \mathcal{G}_k} \left( \frac{1}{2} \left\| \boldsymbol{\beta}_m - \boldsymbol{\beta}_m^{(t)} \right\|_2^2 + \lambda \left\| \boldsymbol{\beta}_m - \boldsymbol{\theta}_k \right\|_2 \right) \right\} \quad \text{ for } k\in [K].\\
\end{aligned}\label{orup}
\end{equation}

\begin{theorem}[Oracle IHT Estimator with Known Group Structure] \label{teoor}
Suppose that Assumptions~1, 2, 3, and 4 hold.
 Let 
\[
\sigma \asymp \left( \frac{v_\delta N_{\min}}{ \log(pM) } \right)^{1/(1+\delta)}, \quad 
\eta = \frac{4}{c_l + 2c_u},\quad \min\left\{c_1,1\right\}\leq \frac{s}{s_0}\leq \frac{q}{q_0} < \infty.
\]
If \( \sqrt{q_0} \left( \frac{\log(pM)}{n} \right)^{\delta/(1+\delta)} = o(1) \), and
\[
\max\left\{ h, e_0, |\mathcal{G}_{\min}|^{1/(1+\delta)} \sqrt{q_0} \left(\frac{\log(pM)}{n}\right)^{\delta/(1+\delta)} \right\} \lesssim \lambda \lesssim \sigma,
\]
then with probability at least \(1 - (pM)^{-c}\) for some positive constant \(c\), and for all \(T \geq 1\), the federated IHT estimators obtained by replacing \eqref{cu} with \eqref{orup} in Algorithm~\ref{fl} with the oracle group structure \( \{ \mathcal{G}_k \}_{k \in [K]} \)—satisfy:
\begin{equation}\label{ore}
\max_{m\in [M]} \left\| \hat{\boldsymbol{\beta}}^{(T)}_m - \boldsymbol{\beta}^*_m \right\|_2 
\lesssim \underbrace{\kappa_1^T}_{\text{iterative error}} + 
\underbrace{ {\sqrt{s_0} \left( \frac{\log(pM)}{N_{\min}} \right)^{\delta/(1+\delta)}} }_{\text{Oracle convergence rate}} + 
\underbrace{h}_{\text{within-group heterogeneity}},
\end{equation}
where \( \kappa_1=\left(1 + \sqrt{\frac{q_0}{q}}\right) \frac{c_u - \frac{c_u}{2}}{c_u + \frac{c_u}{2}} + C \sqrt{\frac{s_0 \log(pM)}{ N_{min}}}<1 \) for some positive constant $C$.
\end{theorem}
\par The proposed estimator achieves a convergence rate that matches the rate attainable by pooling all related data. Since the personalised estimator is $s$-sparse, the convergence rates under both $\ell_2$ and $\ell_1$ norms are accelerated by a factor of $|\mathcal{G}_{\min}|^{1/(1+\delta)}$ compared with local estimators, provided that $h$ is sufficiently small. This theorem improves upon existing results in several respects. When $\delta=1$, $h=0$, and group sizes are balanced, \cite{doi:10.1080/01621459.2024.2321652} achieves the oracle rate 
\(\sqrt{\tfrac{K s_0 \log(pM)}{N_{\min}}}\), which is \(\sqrt{K}\) times larger than ours. \cite{JMLR:v25:23-0059} requires a substantially larger local sample size and a stronger initialization condition to achieve a $\sqrt{|\mathcal{G}_{\min}|}$-fold acceleration over individual local estimators, due to its one-shot estimation framework. In contrast, our IHT-based federated method requires only moderately large local sample sizes and mild initialisation, while applying to a broad range of heterogeneous settings.

\par Next, when the subgroup structure is unknown, we initialise 
\(\hat{z}_m^{(0)} = \tilde{z}_m\), obtained by minimising 
\(\mathcal{L}(\boldsymbol{\beta}, \boldsymbol{\theta}, \mathbf{z};
\{\hat{\boldsymbol{\beta}}_m^{(0)}\}_{m\in[M]})\), 
in a manner similar to~\eqref{cu}.
With appropriately chosen initial estimators 
\(\{\hat{\boldsymbol{\beta}}_m^{(0)}\}_{m\in[M]}\) 
(e.g., the local IHT estimator), 
we establish that Algorithm~\ref{fl} accurately recovers the true group structure 
and that the resulting estimators achieve the same convergence rate 
as in the oracle setting.

\begin{theorem} [Federated IHT Estimator with Unknown Group Structure]
    \label{l4.1}
Suppose that Assumptions~1, 2, 3, 4, and 5 hold. Initialised \(\hat{z}_m^{(0)} = \tilde{z}_m\) and let 
$$
\sigma \asymp \left(v_\delta N_{min} / \log(pM)\right)^{1 /(1+\delta)}, \quad \eta = \frac{4}{c_l + 2c_u},\quad \lambda\asymp\Delta, \quad \min\left\{c_1,1\right\}\leq \frac{s}{s_0}\leq \frac{q}{q_0} <\infty.
$$
If $\sqrt{q_0}\left(\frac{\log (p M)}{n}\right)^{\delta / (1+\delta)}=o(1)$ and {
$
K\max\{h,e_0\}  \lesssim \lambda \lesssim K\sigma,
$}
then with probability at least \(1 - (pM)^{-c}\) for some positive constant \(c\), and for all \(T \geq 1\), the Federated IHT estimators obtained in Algorithm~\ref{fl} satisfy:
\[
\begin{aligned}
\max_{m\in[M]} \left\|\hat{\boldsymbol{\beta}}^{(T)}_m - \boldsymbol{\beta}^*_m\right\|_2 
\lesssim & \underbrace{\kappa_1^{T}}_{\text{iterative error}} + {\underbrace{{\sqrt{s_0}\left(\frac{\log (pM)}{N_{min}}\right)^{\delta / (1+\delta)}}}_{\text{Oracle convergence rate}}}+\underbrace{h}_{\text{within-group heterogeneity}},
\end{aligned}
\]
where \( \kappa_1=\left(1 + \sqrt{\frac{q_0}{q}}\right) \frac{c_u - \frac{c_u}{2}}{c_u + \frac{c_u}{2}} + C \sqrt{\frac{s_0 \log(pM)}{ N_{min}}}<1 \) for some positive constant $C$.
\end{theorem}
Compared with the oracle setting, our method requires the subgroup identifiability condition (Assumption~5) and a stronger initialisation condition, both of which become more demanding as the number of groups $K$ increases. Nevertheless, the initialisation requirement remains mild as long as Assumption~5 is satisfied. Assumption~5 reflects the fact that, as $K$ increases, a stronger level of identifiability is required for clustering, consistent with the assumptions adopted in \cite{10.1214/23-AOS2319,doi:10.1080/01621459.2024.2321652}.
 Regarding sparsity recovery, each task can achieve a result analogous to Corollary~\ref{cor1}, but under a more relaxed condition when $h$ is sufficiently small.

\begin{corollary} \label{cor3}
    Under the assumptions of Theorem~\ref{l4.1}, if
    $
    \sqrt{s_0}\left(\frac{\log (pM)}{N_{\min}}\right)^{\delta / (1+\delta)} + h 
    \ll \beta_{min}.
  $
   then after 
$
T = O\!\left(\log\!\frac{e_0}{\sqrt{s_0} \, (\log(pM)/N_{\min})^{\delta/(1+\delta)}}\right)
$ iterations,
with probability at least \(1 - (pM)^{-c}\) for some constant \(c>0\), the support of 
\(\hat{\boldsymbol{\beta}}_m^{(T)}\) contains that of \(\boldsymbol{\beta}_m^*\), and the number of falsely selected variables is at most \(s - s_0\).
\end{corollary}
\par The following corollary establishes recovery of the latent group structure, extending the clustering guarantees of \cite{10.1214/23-AOS2319} to high-dimensional settings. Moreover, the resulting convergence rate matches the error bound achieved by pooling all related data to estimate the $q_0$-sparse group centre.
\begin{corollary}\label{cor2}
Under the assumptions of Theorem~\ref{l4.1}, with probability at least \(1 - (pM)^{-c}\) for some positive constant \(c\), there exists a permutation \( \tau \) of \([K]\) such that for all \( m \in [M] \), we have:
\begin{equation}
\widehat{z}_m^{(T)} = \tau(z_m^{\star}).
\end{equation}
Furthermore, the following error bound holds:
\[
\max_k \left\|\hat{\boldsymbol{\theta}}_k^{(T)} - \boldsymbol{\theta}_k^*\right\|_2 \lesssim {\kappa_1^T} + {\sqrt{q_0} \left( \frac{\log (pM)}{N_{min}} \right)^{\delta/(1+\delta)}}+h,
\]
where \( \hat{z}_m^{(T)} \) and \( \hat{\boldsymbol{\theta}}_{k}^{(T)} \) are obtained from Algorithm~\ref{fl}.
\end{corollary}

\par
\begin{figure}[ht]
  \centering
  \includegraphics[width=0.45\textwidth, height=0.2\textheight]{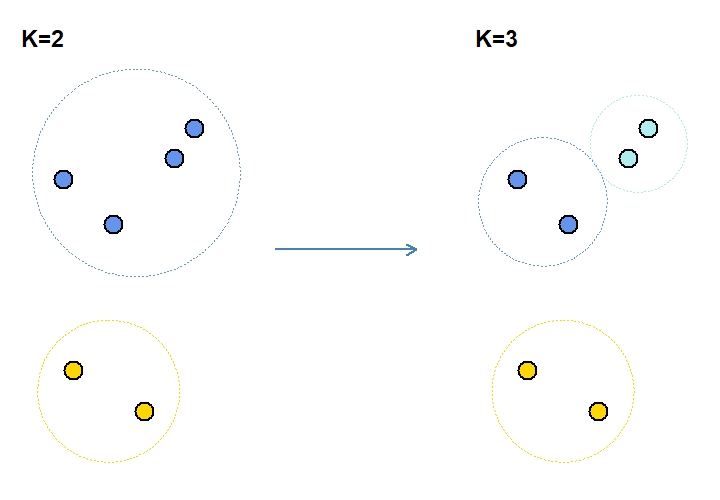}
\caption{Given $\{\boldsymbol{\beta}_m^*\}_{m\in[M]}$, different choices of $K$ lead to distinct subgroup structures with varying $h$, $\Delta$, and group sizes.}
  \label{KMK}
\end{figure}

 Adaptively tuning the number of groups helps alleviate the negative impact of excessive within-group heterogeneity. Theorems above show that large $h$ can make federated estimation less efficient than local estimation. Simulations in Supplementary Section~C.5 show that, in the low-dimensional case, high heterogeneity affects our method more than \cite{10.1214/23-AOS2319}. This reflects the inherent cost of our method—requiring a sufficiently large $\lambda$ for convergence and accurate clustering in distributed optimisation, thereby raising the heterogeneity cost to $h$.  One way to mitigate it is to increase the number of groups by assigning tasks that are far from their group centre to new groups. While this decreases within-group heterogeneity, it may lead to smaller group sizes and reduced separation $\Delta$. Figure~\ref{KMK} illustrates that, given the true task parameters, different choices of the number of groups can yield distinct structures, each characterised by a unique combination of within-group heterogeneity, between-group separation, and group size. There can be multiple values for $K$ that produce subgroup structures which meet our identifiability requirements. Therefore, adaptively selecting the number of groups $K$ in practice is crucial for balancing these factors. Our subsequent simulation results corroborate this finding.
\par Finally, although our analysis assumes common sparsity $s_0$ and balanced sample sizes $n$ across tasks for simplicity, the theoretical results can be extended to more general settings. Specifically, the IHT method can accommodate task-specific sparsity levels \( s_m \), allowing each estimator to vary in sparsity. The same convergence rate holds as long as the assumptions are satisfied for each \( s_m \). Furthermore, with varying sample sizes \( n_m \) across tasks, a near-optimal convergence rate requires $\max_m \{n_m\} = O(\min_m \{n_m\})$.

\section{Simulation \label{simu}}
This section presents simulation results to empirically validate our theoretical insights. We consider $M$ tasks, where each task generates data from the linear model:
\[
y_{mi} = \mathbf{x}_{mi}^T \boldsymbol{\beta}_{m} + \epsilon_{mi}.
\]
The design matrix $\mathbf{x}_j$ is generated from a multivariate normal distribution with mean 0 and $\operatorname{cov}(x_{is}, x_{it}) = 0.3$ for $s \neq t$ and $\operatorname{cov}(x_{it}, x_{it}) = 1$ for $s, t \in [p]$. 
\par {We conduct four sets of experiments to evaluate the proposed algorithm:  
(a) The effect of Clustered Federated Learning when both inter-group and within-group heterogeneity are present, as well as the influence of the number of tasks. (b) The influence of heavy-tailed noise. (c) The influence of within-group heterogeneity. (d) The influence of signal-to-noise ratio (SNR) and inter-group heterogeneity.
In summary, our experimental investigation considers the following parameter configurations}:
\begin{itemize}
    \item \textbf{Setting 1 (Clustered Federated Learning):}  
    \[
    \boldsymbol{\beta}_m = \boldsymbol{\theta}_{z_m} + \boldsymbol{h}_m,
    \]  
    where group centres are \(\boldsymbol{\theta}_1 = (2, 3, 4, 0, \dots, 0)\) and \(\boldsymbol{\theta}_2 = (-1, 2, 3, 0, \dots, 0)\).  
    Here, \(\boldsymbol{h}_m = (\boldsymbol{h}_{m1}, \mathbf{0})\) with \(\boldsymbol{h}_{m1} \sim N(\mathbf{0}_3, 0.3^2 \boldsymbol{I}_{3})\).  
    Tasks are divided into two groups: \(z_m = 1\) for \(m \in \{1, \dots, 0.6M\}\) and \(z_m = 2\) for \(m \in \{0.6M+1, \dots, M\}\).  
    The errors follow a \(t(2)\) distribution.  

    \item \textbf{Setting 2 (Noise Types):}  
    Same as Setting~1, except \(\boldsymbol{h}_{m1} \sim N(\mathbf{0}_3, 0.1^2 \boldsymbol{I}_{3})\).  
    We consider three noise distributions: Gaussian \(N(0,1)\), \(t(2)\), and Cauchy \(C(1.5)\).  

    \item \textbf{Setting 3 (Within-Group Heterogeneity):}  
    Same as Setting~1, but
    \[
    \boldsymbol{\beta}_m = \boldsymbol{\theta}_{z_m} + h \cdot \frac{\boldsymbol{h}_m}{\|\boldsymbol{h}_m\|_2},
    \]  
    where \(h\) controls the degree of within-group heterogeneity.  

    \item \textbf{Setting 4 (SNR and Between-Group Heterogeneity):}  
    Same as Setting~1, but
    \[
    \boldsymbol{\beta}_m = \Delta \cdot \boldsymbol{\theta}_{z_m} + \boldsymbol{h}_m,
    \]  
    where \(\Delta\) simultaneously controls the signal-to-noise ratio (SNR) and the degree of between-group heterogeneity.  
\end{itemize}
For each simulation setting, we generate 100 datasets and evaluate the methods using the following metrics: (a)  Mean Squared Error (MSE) of the estimated parameters, computed as \(\frac{1}{M} \sum_{m=1}^M \| \hat{\boldsymbol{\beta}}_m^{(T)} - \boldsymbol{\beta}_m^* \|_2^2\); (b) The average numbers of false negatives (FN) and false positives (FP) for each task for selected variables, and (c) Rand Index (RI), measuring the similarity between estimated and true cluster labels, with higher values indicating better performance.

\par We compare the performance of the following methods:
\begin{itemize}
    \item \texttt{IHT-local}: Local estimators obtained from Algorithm \ref{localIHt} for each task independently.
    \item \texttt{IHT-GP}: Federated estimators obtained from Algorithm \ref{fl}.
   \item \texttt{PerFL}: Federated estimators with SCAD regularizers introduced in \cite{doi:10.1080/01621459.2024.2321652}.

\item \texttt{IHT-ML}: Pooling all data to optimize
\begin{equation}
\begin{aligned}
    \underset{
        \substack{
            \boldsymbol{\beta} \in \mathbb{R}^{p \times M}, \;
            \boldsymbol{\theta} \in \mathbb{R}^{p \times K}, \;
            \mathbf{z} \in [K]^M
        }
    }{\arg\min} \; 
    \sum_{m=1}^M H_\sigma^m(\boldsymbol{\beta}_m) 
    + \lambda \left\|\boldsymbol{\theta}_{z_m} - \boldsymbol{\beta}_m\right\|_2, \\
    \text{subject to } \quad 
    \|\boldsymbol{\beta}_m\|_0 \leq s, \;
    \|\boldsymbol{\theta}_k\|_0 \leq q, 
    \quad m \in [M], \; k \in [K].
    \label{lss}
\end{aligned}
\end{equation}

The optimization procedure requires access to all data and follows a combination of the ML algorithm in \cite{10.1214/23-AOS2319} and the IHT method. The implementation details are presented in Supplementary Section~C.1.
\item \texttt{IHT-L2}: Federated estimators obtained by replacing the Huber loss with the \(L_2\) loss in Algorithm \ref{fl}.
\item \texttt{Oracle}: Federated estimators obtained using the true class labels $\{z_m\}$ in Algorithm~\ref{fl}.
\end{itemize}
For initialization, \texttt{IHT-local} is initialized with estimators obtained from penalized Huber regression, as proposed in \cite{doi:10.1080/01621459.2018.1543124}, and all federated methods adopt \texttt{IHT-local} as the initial estimate. For parameter selection, we fix \(\sigma = 3\) and \(\eta = 0.01\), proximal map step size $\eta_1=0.01$. Because $q_0 = s_0$ in these simulation settings, we set $q = s$ for simplicity. We specify a grid of values for $(K, s, \lambda)$ and select the optimal parameters by minimizing the following objective:
\[
\frac{1}{Mn}\sum_{m=1}^M \sum_{i=1}^{n} H_\sigma\!\left(y_{mi} - \mathbf{x}_{mi}^{\top} \widehat{\boldsymbol{\beta}}_m\right) 
+ \frac{\log p}{n}\,(s + 1.5K).
\]
When $q_0 > s_0$, we apply the same criterion with an additional grid search over $q$. For \texttt{PerFL}, the regularization parameter is selected via the Bayesian Information Criterion (BIC) as recommended in \cite{doi:10.1080/01621459.2024.2321652}. While \texttt{PerFL} adaptively clusters tasks based on feature-wise similarity, it does not return explicit task labels. For evaluation, we therefore apply $k$-means clustering (with $K=2$) to the \texttt{PerFL} estimates and assess clustering accuracy.

\begin{remark}
The \texttt{IHT-ML} method is a baseline ML estimator with full data access and no privacy guarantees. While it performs competitively in our experiments, a rigorous theoretical analysis is left for future work. 
\end{remark}
\subsection{Simulation results}
\par {Table~\ref{tab:t1}} presents the numerical results for Setting~1. Across scenarios, the proposed \texttt{IHT-GP} estimators achieve results closest to those of the Oracle and ML estimators, effectively recovering the subgroup structure and achieving estimation performance comparable to pooling all data. The \texttt{PerFL} method performs poorly because it lacks explicit modeling of within-group heterogeneity and requires a stricter identifiability condition for clustering, which may fail when ($n/p$) is small. Additional comparisons for larger $n/p$ and varying $M$ are provided in Supplementary Section~C.6 and Section~C.7, which report similar findings.
\begin{table}[!ht]
  \centering
\caption{Performance comparison under Setting 1 for various $(n,p)$ with $M=10$.}
  \resizebox{\textwidth}{!}{%
  \begin{threeparttable}
    \begin{tabular}{lcccccccccccccccc}
      \hline
      (n, p) & \multicolumn{4}{c}{(50, 100)} & \hspace{4pt} & \multicolumn{4}{c}{(50, 300)} & \hspace{4pt} & \multicolumn{4}{c}{(50, 500)} \\
      \hline
      Method & MSE & FP & FN & RI & & MSE & FP & FN & RI & & MSE & FP & FN & RI \\
      \hline
      \texttt{IHT-local} & 1.518 & 0.450 & 0.390 & -     & & 2.727 & 0.612 & 0.552 & -     & & 3.136 & 0.618 & 0.558 & -     \\
      \texttt{IHT-GP}    & 0.362 & 0.088 & 0.088 & 1.000 & & 0.392 & 0.104 & 0.104 & 1.000 & & 0.446 & 0.158 & 0.138 & 0.992 \\
      \texttt{PerFL}     & 0.836 & 0.132 & 0.078 & 0.699 & & 1.179 & 0.138 & 0.242 & 0.422 & & 1.376 & 0.592 & 0.422 & 0.242 \\
      \texttt{IHT-ML}    & 0.373 & 0.044 & 0.044 & 0.998 & & 0.388 & 0.068 & 0.068 & 0.992 & & 0.439 & 0.090 & 0.090 & 0.997 \\
      \texttt{IHT-L2}    & 2.314 & 0.258 & 0.198 & 0.768 & & 2.803 & 0.556 & 0.356 & 0.636 & & 2.928 & 0.516 & 0.336 & 0.640 \\
      \texttt{Oracle}    & 0.350 & 0.002 & 0.002 & 1.000 & & 0.385 & 0.218 & 0.138 & 1.000 & & 0.440 & 0.344 & 0.204 & 1.000 \\
      \hline
      (n, p) & \multicolumn{4}{c}{(100, 100)} & \hspace{4pt} & \multicolumn{4}{c}{(100, 300)} & \hspace{4pt} & \multicolumn{4}{c}{(100, 500)} \\
      \hline
      Method & MSE & FP & FN & RI & & MSE & FP & FN & RI & & MSE & FP & FN & RI \\
      \hline
      \texttt{IHT-local} & 0.547 & 0.188 & 0.168 & -     & & 0.702 & 0.280 & 0.240 & -     & & 0.844 & 0.290 & 0.272 & -     \\
      \texttt{IHT-GP}    & 0.248 & 0.008 & 0.008 & 1.000 & & 0.250 & 0.016 & 0.016 & 1.000 & & 0.254 & 0.008 & 0.008 & 1.000 \\
      \texttt{PerFL}     & 0.495 & 0.002 & 0.004 & 0.976 & & 0.522 & 0.002 & 0.602 & 0.913 & & 0.704 & 1.578 & 0.426 & 0.851 \\
      \texttt{IHT-ML}    & 0.246 & 0.004 & 0.004 & 1.000 & & 0.255 & 0.006 & 0.006 & 1.000 & & 0.259 & 0.006 & 0.006 & 1.000 \\
      \texttt{IHT-L2}    & 0.651 & 0.188 & 0.088 & 0.888 & & 0.839 & 0.056 & 0.056 & 0.859 & & 1.923 & 0.326 & 0.126 & 0.843 \\
      \texttt{Oracle}    & 0.231 & 0.052 & 0.012 & 1.000 & & 0.248 & 0.068 & 0.008 & 1.000 & & 0.251 & 0.130 & 0.030 & 1.000 \\
      \hline
    \end{tabular}
  \end{threeparttable}%
  }
  \label{tab:t1}
\end{table}

\begin{table}[!ht]
  \centering
  \caption{Performance comparison across different noise types under Setting 2, with \(n = 100\), \(p = 300\), and \(M = 10\).}
  \resizebox{\textwidth}{!}{%
    \begin{threeparttable}
      \begin{tabular}{lcccc ccccc ccccc}
        \hline
        & \multicolumn{4}{c}{Normal} & & \multicolumn{4}{c}{t} & & \multicolumn{4}{c}{Cauchy} \\
        \hline
        Method & MSE & FP & FN & RI & & MSE & FP & FN & RI & & MSE & FP & FN & RI \\
        \hline
        \texttt{IHT-GP}   & 0.031 & 0.000 & 0.000 & 1.000 & & 0.101 & 0.120 & 0.020 & 1.000 & & 1.269 & 1.365 & 0.265 & 0.765 \\ 
        \texttt{PerFL}    & 0.031 & 0.000 & 0.000 & 1.000 & & 0.361 & 0.210 & 0.008 & 0.940 & & 3.472 & 7.250 & 0.575 & 0.272 \\ 
        \texttt{IHT-ML}   & 0.030 & 0.000 & 0.000 & 1.000 & & 0.097 & 0.065 & 0.015 & 0.995 & & 1.132 & 0.713 & 0.538 & 0.776 \\ 
        \texttt{IHT-L2}   & 0.030 & 0.003 & 0.003 & 1.000 & & 1.164 & 0.390 & 0.115 & 0.931 & & 18.552 & 1.793 & 1.393 & 0.035 \\ 
        \texttt{Oracle}   & 0.029 & 0.000 & 0.000 & 1.000 & & 0.102 & 0.008 & 0.000 & 1.000 & & 0.320 & 0.210 & 0.010 & 1.000 \\ 
        \hline
      \end{tabular}
    \end{threeparttable}%
  }
  \label{tab:t2}
\end{table}

\par Table~\ref{tab:t2} presents our numerical results on robustness to heavy-tailed noise, highlighting the advantages of the Huber loss. Our \texttt{IHT-GP} method achieves competitive performance across different noise distributions. For the Normal distribution, an appropriate choice of $\sigma$ yields efficiency comparable to the squared loss. Under the Student's $t$ and Cauchy distributions, the Huber loss performs markedly better than the squared loss. Additional simulations in Supplementary Section~C.8 exhibit similar patterns when varying the degrees of freedom of the Student’s $t$, further confirming the robustness of our method to heavy-tailed noise.

\begin{figure}[ht!]
  \centering
  \includegraphics[width=0.8\textwidth, height=0.35\textheight]{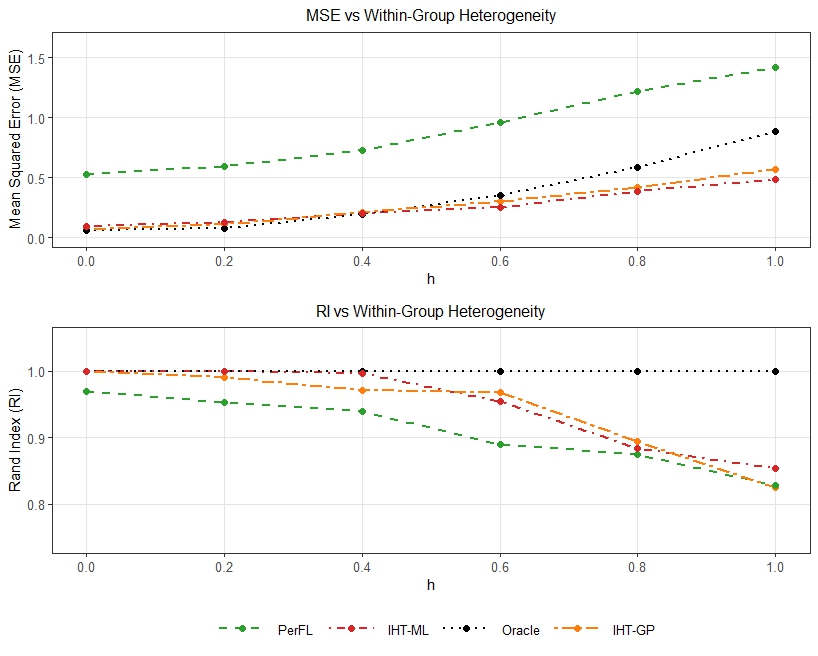}
 \caption{Influence of within-group heterogeneity on MSE and RI under Setting~3, with \(n=100\), \(p=300\), \(M=10\).}
    \label{fig:f2}
\end{figure}
\par {Figures~\ref{fig:f2}} illustrate the effects of within-group heterogeneity under Setting~3. As \(h\) increases, both the MSE and RI gradually deteriorate for all clustering methods except the RI of the oracle one. However, when \(h\) becomes sufficiently large, the MSE of the oracle method is no longer optimal, indicating that the true clustering structure may become suboptimal. In such cases, adaptively tuning \(K\) to account for latent heterogeneity is more appropriate, as evidenced by the lower MSE of \texttt{IHT-GP}. Consequently, the RI may decrease, since the default true grouping assumes only two clusters. This finding corroborates our discussion in the previous section, indicating that adaptively tuning \(K\) can reduce within-group heterogeneity \(h\) and thereby mitigate its adverse effects.
\begin{figure}[ht!]
  \centering
  \includegraphics[width=0.8\textwidth, height=0.35\textheight]{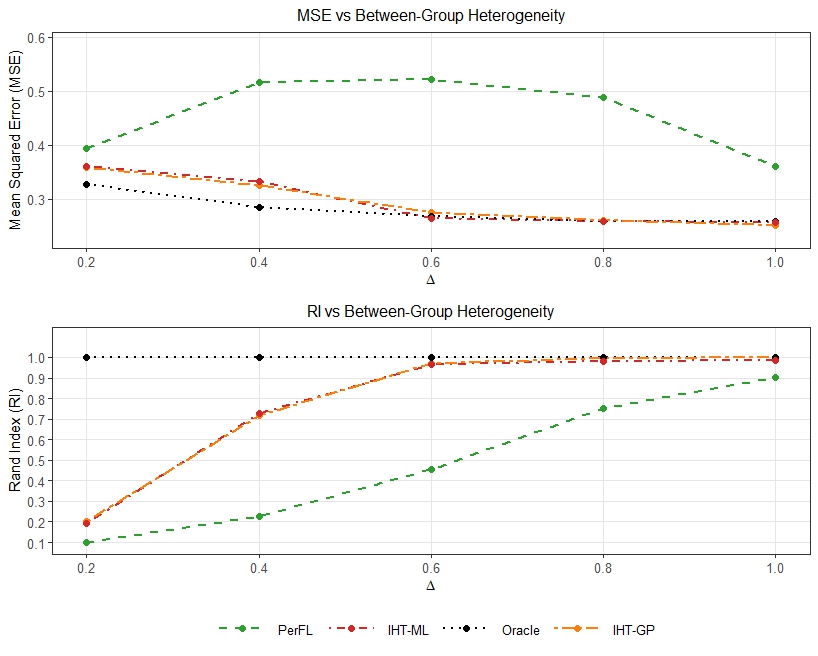}
  \caption{{Influence of SNR and between-group heterogeneity on MSE and RI under Setting 4, with \(n = 100\), \(p = 300\), and \(M = 10\)}.}
  \label{fig:f3}
\end{figure}
\par {Figure~\ref{fig:f3}} illustrates the effects of different levels of inter-group heterogeneity under Setting 4. As \( \Delta \) increases, both clustering accuracy and estimation efficiency improve. In this regime, our proposed methods demonstrate advantages in both estimation accuracy and clustering performance, requiring only a relatively large \( \Delta \). 

\section{Real-world Data \label{s6}}
We assess the effectiveness of the proposed algorithms using the Genotype-Tissue Expression (GTEx) dataset (\url{https://gtexportal.org/}), 
which records gene expression from 49 tissues of 838 individuals, comprising 1,207,976 observations on 38,187 genes. 
Our analysis focuses on MODULE\_137, consisting of 545 core genes and 1,632 additional genes significantly enriched under the same conditions. 
The number of observations per tissue ranges from 85 to 803.  

We aim to predict the expression level of gene DVL1 (response variable) and uncover the latent group structure across tissues. 
For each tissue, the data are split into five folds, with one fold used for testing and the remaining four for training. 
Method performance is evaluated by: (i) {PE}, the average prediction error on test data across tissues; 
(ii){SE}, the corresponding standard error; and (iii) {Size}, the average number of selected variables, defined as
$
\frac{1}{49}\sum_{m=1}^{49}\|\hat{\boldsymbol{\beta}}_m\|_0.$
All results are averaged over 50 independent replications and compared across the following three methods:

\begin{itemize}
\item \texttt{Huber-lasso}: Applies the penalized Huber regression method of \cite{doi:10.1080/01621459.2018.1543124} to each tissue independently.
\item \texttt{Pooling}: Combines data from all tissues to fit a global model by \texttt{Huber-lasso}.
\item \texttt{IHT-GP}: Applies Algorithm~\ref{fl} to these tissues.
\end{itemize}
\
\begin{table}[htbp]
\centering
\caption{ Performance comparison of different methods for MODULE\_137 dataset.}
\label{tab:results2}
{\fontsize{12}{12}\selectfont
\begin{tabular}{lccc}
\toprule
Method & \multicolumn{1}{c}{PE} & \multicolumn{1}{c}{SE} & \multicolumn{1}{c}{Size} \\
\midrule
\texttt{Huber-lasso} & 0.195
 & 0.031 & 605.18 \\
\texttt{Pooling}     & 0.236
 & 0.007 & 72.26 \\
\texttt{IHT-GP}      & 0.158
 & 0.028 & 22.36
 \\
\hline
\end{tabular}
}
\end{table}
The results are summarized in Table~\ref{tab:results2}. 
The \texttt{Huber-lasso}, which relies solely on local information, serves as the baseline. 
Owing to the limited sample sizes in some tissues, it tends to select an excessively large number of variables. 
The pooling estimator performs the worst in prediction, as it ignores heterogeneity across tissues. 
In contrast, our proposed estimators achieve the lowest prediction errors and yield substantially more parsimonious models by adaptively clustering tissues and sharing information.
\begin{figure}[ht!]
  \centering
  \includegraphics[width=0.75\textwidth, height=0.25\textheight]{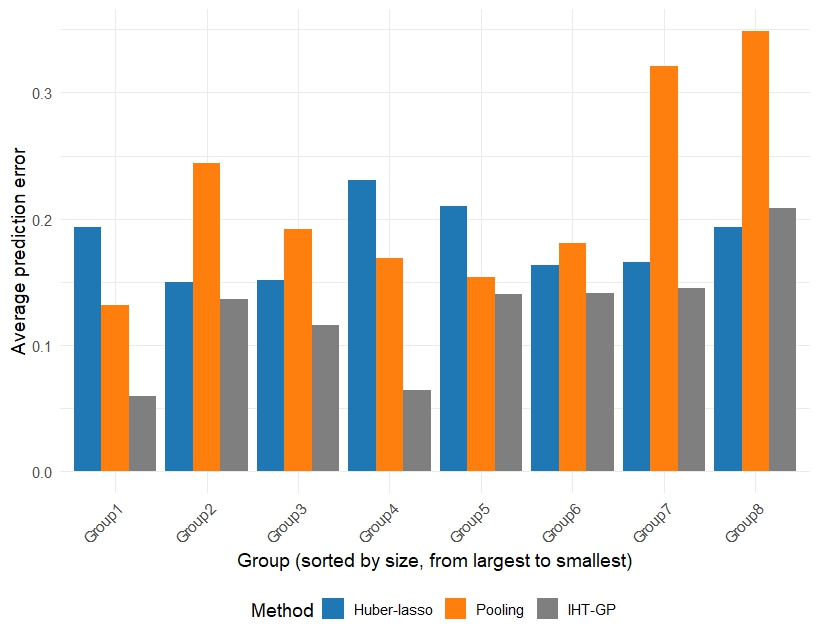}
\caption{Group-level average prediction errors of the three methods for gene {DVL1} across 8 identified nontrivial clusters.}
  \label{fig:DVL}
\end{figure}
\par Regarding the latent group structure, we identify 8 nontrivial clusters (sizes 2–6) across the 49 tissues (see Supplementary Section~C.9). 
The clustering is imbalanced, reflecting underlying heterogeneity. 
Ordered by group size, Groups~1, 3, 5, and 6 each consist of tissues from a single system (Nervous, Digestive, Nervous, and Skin/Glands, respectively), while Group~4 is dominated by Nervous tissues. 
Group~2 spans multiple systems (Immune, Digestive, Endocrine, Reproductive, and Skin/Glands), suggesting functional interplay, and the remaining two groups are also mixed. 
Overall, the subgroups broadly align with human physiological systems and functional interactions. 
Figure~\ref{fig:DVL} reports the average PE within each group for the three methods, showing that \texttt{IHT-GP} consistently outperforms the others in most groups, underscoring the effectiveness of our clustering approach.

\section{Conclusion} \label{s7}
\par In this paper, we propose a robust personalized federated learning framework for heterogeneous multi-source data. Our approach improves estimation efficiency and accommodates heterogeneity through task clustering while preserving privacy. By integrating IHT with Huber regression, it achieves robustness to heavy-tailed noise and strong variable selection performance. Theoretically, we establish linear convergence and derive non-asymptotic error bounds, and both simulations and real data confirm its superiority over existing methods.

\par Several avenues for future research remain. 
First, our heterogeneity assumption may be restrictive in terms of local sample sizes and inter-group separability $\Delta$; relaxing these conditions while maintaining theoretical guarantees is important. 
Second, a deeper analysis of the solution to \eqref{lss} would strengthen its applicability in high-dimensional multitask learning. 
Third, since $k$-means is sensitive to initialization and requires pre-specifying $K$, developing more robust clustering alternatives is a promising direction.

\bibliographystyle{plainnat} 
\bibliography{references}   

\end{document}